\documentclass[runningheads]{llncs}
\usepackage{graphicx}
\usepackage{amsfonts}
\usepackage{amsmath} 
\usepackage{array}
\usepackage{multirow}
\usepackage{marvosym}
\usepackage{float}
\newcommand{\tabincell}[2]{\begin{tabular}{@{}#1@{}}#2\end{tabular}}
\usepackage{setspace}

\usepackage[colorlinks, linkcolor=blue, anchorcolor=blue, urlcolor=blue, citecolor=blue]{hyperref}
\def\mathbi#1{\textbf{\em #1}}
\begin{document}
\title{CSSR: A Context-Aware Sequential Software Service Recommendation Model \thanks{Please cite the paper as the following: Zhang, M., Liu, J., Zhang, W., Deng, K., Dong, H., Liu, Y.: CSSR: A Context-Aware Sequential Software Service Recommendation Model. The 19th International Conference on Service Oriented Computing (ICSOC 2021), 2021. \url{https://doi.org/10.1007/978-3-030-91431-8_45}.}}
\titlerunning{Context-Aware Sequential Software Service Recommendation}

\author{Mingwei Zhang\inst{1} \and Jiayuan Liu\inst{1} \and Weipu Zhang\inst{1} \and Ke Deng\inst{2}\textsuperscript{(\Letter)} \and Hai Dong\inst{2} \and Ying Liu\inst{1}}
\authorrunning{M. Zhang et al.}
\institute{Software College, Northeastern University, Shenyang, China\\
	\email{\{zhangmw,liuy\}@swc.neu.edu.cn}, \email{\{2071295,1871164\}@stu.neu.edu.cn}\and
	School of Computing Technologies, RMIT University, Melbourne, Australia\\
	\email{\{Ke.Deng,Hai.Dong\}@rmit.edu.au}}
\maketitle              % typeset the header of the contribution

\begin{abstract}
We propose a novel software service recommendation model to help users find their suitable repositories in GitHub. Our model first designs a novel context-induced repository graph embedding method to leverage rich contextual information of repositories to alleviate the difficulties caused by the data sparsity issue. It then leverages sequence information of user-repository interactions for the first time in the software service recommendation field. Specifically, a deep-learning based sequential recommendation technique is adopted to capture the dynamics of user preferences. Comprehensive experiments have been conducted on a large dataset collected from GitHub against a list of existing methods. The results illustrate the superiority of our method in various aspects.  


\end{abstract}
\section{Introduction}
With the development of emerging computing areas such as cloud computing, big data, and Internet of Things, the Web-based services available on the Internet have increased rapidly in both quantity and type \cite{Overview1}. Following \cite{PNCF}, software service is specifically defined as \emph{services which contain code under open-source licenses for others to use and modify freely}, such as open-source projects or repositories on social coding sites (e.g., GitHub\footnote[1]{https://github.com/}, Bitbucket\footnote[2]{https://bitbucket.org/}, SourceForge\footnote[3]{https://sourceforge.net/}).  Users can build their Web services, applications, or even scientific experiment systems quickly by exploiting functional code modules in massive software services \cite{paper2repo}.

As a representative software service hosting platform,  GitHub is widely known to developers from all over the world, who find it easier and quicker to build up their complex applications from particular repositories.  As of January 2020, GitHub reports having over 40 million users and more than 100 million repositories \cite{GitHub1},  making it the largest host of software services in the world. The large number of repositories has undoubtedly increased the difficulty of selecting the most suitable ones to fulfill users' application development. Therefore, software service recommendation has become of practical importance \cite{GHTRec}. McMillan et al. \cite{CLAN} designed a tool named CLAN to help users detect similar applications. It used latent semantic indexing to measure the similarity of repositories relative to API usage. However, it is limited to Java applications and small-scale data. LRMF\cite{LRMF} is a pairwise regularization framework for GitHub open source repository recommendation based on matrix factorization, focusing mainly on exploiting user language preference. PNCF \cite{PNCF} is the state-of-the-art repository recommender model, which combined deep learning with collaborative filtering to enhance recommendation effectiveness, and also focused on language preference.

The above methods proposed effective strategies to make software service recommendation. However, they suffer from the following two common issues. First, \textit{the well-known data sparsity problem is not addressed}. Although the number of users and repositories on GitHub can be very large, the interactions between users and repositories are highly sparse, i.e., most users typically interact with a few repositories. Second, \textit{user preferences may exhibit dynamic characteristics}. For instance, users' preferences may drift over time due to the continuous evolution of software technology and the influence of other users. 

\begin{figure}
	\centering
	\includegraphics[width=0.75\textwidth]{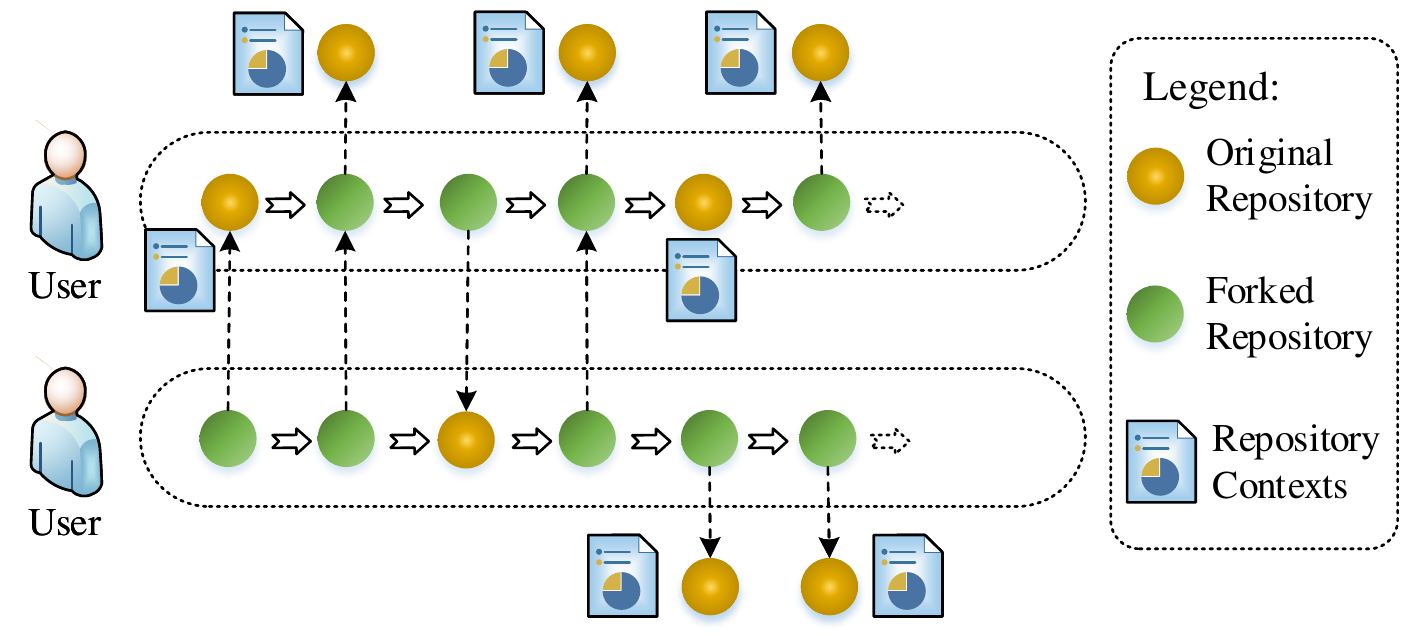}
	\caption{Organization of users and repositories on GitHub.} \label{GitHub}
\end{figure}

To address the two aforementioned issues, a novel recommendation model named CSSR (\underline{C}ontext-aware \underline{S}equential Software \underline{S}ervice \underline{R}ecommendation) is proposed in this paper with two unique traits. Firstly, there is a consensus that contextual data can be utilized as complementary information to alleviate the data sparsity issue in general recommender systems \cite{Rec/Context}.  Each original repository on GitHub has rich contexts as depicted in Fig.~\ref{GitHub}.  We leverage more comprehensive contextual information of repositories (i.e., topics, general description, \textit{R\!E\!A\!D\!M\!E})  compared with the state-of-the-art recommendation methods \cite{PNCF,LRMF}. It can model the similarity between repositories more precisely to make better recommendation when the interaction data is sparse. Secondly, users on GitHub interact with repositories in a chronological order as depicted in Fig.~\ref{GitHub}. The temporal information of user interaction behaviors can help to model users' dynamic interests. For example, it is reasonable to assume that a user is most likely to access the repositories which are relevant to the repositories the user has interacted with recently. Therefore, we adopt sequential user-repository interactions to capture the dynamics of user preferences. More specifically, CSSR first explores contextual information to construct a repository graph upon which the latent vector of each repository can be derived through the graph embedding. Then, the repository sequences of users are fed into a GRU model, where the latent vector of each repository is applied, to identify the appropriate repositories and recommend them to users. We have conducted comprehensive experiments to compare CSSR with the state-of-the-art methods on a large real-world dataset crawled from GitHub\footnote[4]{The dataset and source code are released on https://github.com/JiaYuan6/CSSR.}. The experimental results show that CSSR achieves at least 16.16\%, 22.05\% and 11.35\% improvements over the best baseline in terms of Hit Rate, Mean Reciprocal Rank and Normalized Discounted Cumulative Gain respectively, and the performance boost is more significant in the situation of high level of data sparsity. Our contributions are summarized as follows:

\begin{itemize}
	\item[\textbf{–}] This study proposes a context-aware sequential software service recommendation model (CSSR) to improve software service recommendation by  mitigating the difficulties caused by data sparsity and the dynamic characteristics of user preferences.
	\item[\textbf{–}] We design a repository graph construction method. It 	takes full advantages of the context characteristics of 	repositories on GitHub, upon which the recommendation performance can be improved. 
	\item[\textbf{–}] This study, to the best of our knowledge, is the first work that explores the sequential interactions in software service recommendation to model the dynamics of user preferences. 	
\end{itemize}
The rest of paper is organized as follows: Sect.\ref{sec/RW} introduces the related work; Sect.\ref{sec/PF} formulates the software service recommendation problem; Sect.\ref{sec/Methodology} presents the CSSR model; Sect.\ref{sec/experiments} is the experiments and result analysis, and Sect.\ref{sec/conclusion} is the conclusion.

\section{Related Work}\label{sec/RW}
In this section, we will briefly review several lines of works closely related to ours, including sequential recommendation, service recommendation and software service recommendation.

\subsubsection{Sequential Recommendation.}
The technology of recommender systems continues to develop. This results from their significant role in helping users alleviate the problem of information explosion and select interesting contents in many Web application domains. Meanwhile, many recommendation techniques have been designed, including collaborative filtering\cite{Item_KNN}, matrix factorization\cite{BPR},  factorization machine\cite{FFM}, and deep-learning-based methods\cite{Rec/Context}. Sequential recommendation is a critical research topic of recommender systems, and has been extensively studied over years. It views the interactions as a sequence in a time order and aims to predict the successive items that a user is likely to interact with in the near future. Traditional  methods \cite{SequenceR1} utilized Markov Chains to capture item-item transitions for sequential recommendation. Recently, deep-sequential-neural-model-based methods \cite{GRU4Rec,Bert4Rec}  have shown much superior performance. They utilized different sequential neural networks (e.g., RNNs, Transformers) to address concrete issues from different aspects and domains. Aiming at validating the effectiveness of sequential recommendation in software service domain, we adopt the classic sequential neural model GRU \cite{GRU}  fusing with  graph embedding techniques to solve next repository recommendation issue.

\subsubsection{Service Recommendation.}
Lots of research works have been done on service recommendation in recent years. They can be mainly classified into two categories, i.e., QoS-based service recommendation\cite{QoSR2} and functionality-based service recommendation\cite{SemanticR1,SocialR1,DeepR1}. However, QoS-based service recommendation approaches cannot help developers find an unknown but interesting service, limited by their objective. Functionality-based service recommendation approaches focus on finding the services that meet the functional requirements the best. Among them, the semantic-based approaches\cite{SemanticR1} aim at finding services with the highest matching degree  via semantic similarity computation. The social-network-based approaches\cite{SocialR1} tend to apply user interest, social relationship and link prediction. The information-network-based approaches\cite{DeepR1} mainly employ different kinds of information and multiple semantic meanings of meta paths to recommend services. In this paper, we focus on recommender systems for a special kind of Web-based services, i.e., software services.

\subsubsection{Software Service Recommendation.}
Software services are the services with a focus on providing various code resources to facilitate software development. GitHub is the largest software service providers. It opens access to the information of repository, users, and the interactions between them. This attracts much interest of researchers, and currently most studies in software service recommendation are for GitHub repositories. Jiang et al. \cite{LRMF} proposed a repository recommendation model based on matrix factorization. Chen et al. \cite{PNCF} combined deep learning with collaborative filtering to do repository recommendation. The above two methods both concentrated on exploiting user language preference. Sun et al.\cite{RepoR1} proposed an approach to recommend repositories considering both user behaviors and repository features. Shao et al.\cite{paper2repo} designed a novel cross-platform recommender system, \textit{paper2repo}. It recommended relevant repositories on GitHub that match a given paper, by integrating text encoding and constrained graph convolutional networks.  

In summary, the study in software service recommendation is still in its early stage. None of the existing studies pay attention to making performance analysis in different sparsity levels and addressing the really severe data sparsity issue, together with the problem of dynamic characteristics of user preferences. This paper aims to fill the research gap. 
%have fully addressed the data sparsity issue

\section{Problem Formulation}\label{sec/PF}
Each user has interacted with a sequence of repositories ordered by time on GitHub. The repositories may be created by a user directly or forked from other users, as illustrated in Fig.~\ref{GitHub}. Each repository has contextual information.

Let $\mathcal{U}=\{u_1, u_2, ..., u_{|\mathcal{U}|}\}$ and $\mathcal{R}=\{r_1, r_2, ..., r_{|\mathcal{R}|}\}$ be sets of users and repositories, with $|\mathcal{U}|$ and $|\mathcal{R}|$ being the sizes, respectively.  Each user $u$ can be associated with a sequence of repositories $\mathcal{R}^u=\{r_1^u, r_2^u, ..., r_{|\mathcal{R}^u|}^u\}$ by sorting interaction records in a chronological order, where $r_t^u$ represents the repository that user $u$ interacted with at time step $t$. $\mathcal{R}_{t_1:t_2}^u (t_1<t_2)$ refers to the subsequence from interaction $r_{t_1}^u$ to $r_{t_2}^u$. We learn how to recommend the next repository for each user based on the recent repository subsequence of length $L$ the user interacted with, where $L$ is a hyperparameter. For each user $u$ at time step $t$, we will have a training data record where the features are $R^u_{t-L:t-1}$ and the label is $r^u_t$. For all users from $t=1$ to $t_{cur}-1$ ($t_{cur}$ denotes the current time step and $t_{cur}-1$ is greater than $L$), we will obtain a training data set. Based on it, the research problem investigated in this study is to (1) represent user preferences and repositories, and  (2) develop a prediction model to identify and recommend the preferable repositories to users. The objective is to optimize the performance by addressing the data sparsity issue and the dynamics of user preferences along with time. 

\section{Methodology}\label{sec/Methodology}
\begin{figure}
	\centering
	\includegraphics[width=0.85\textwidth]{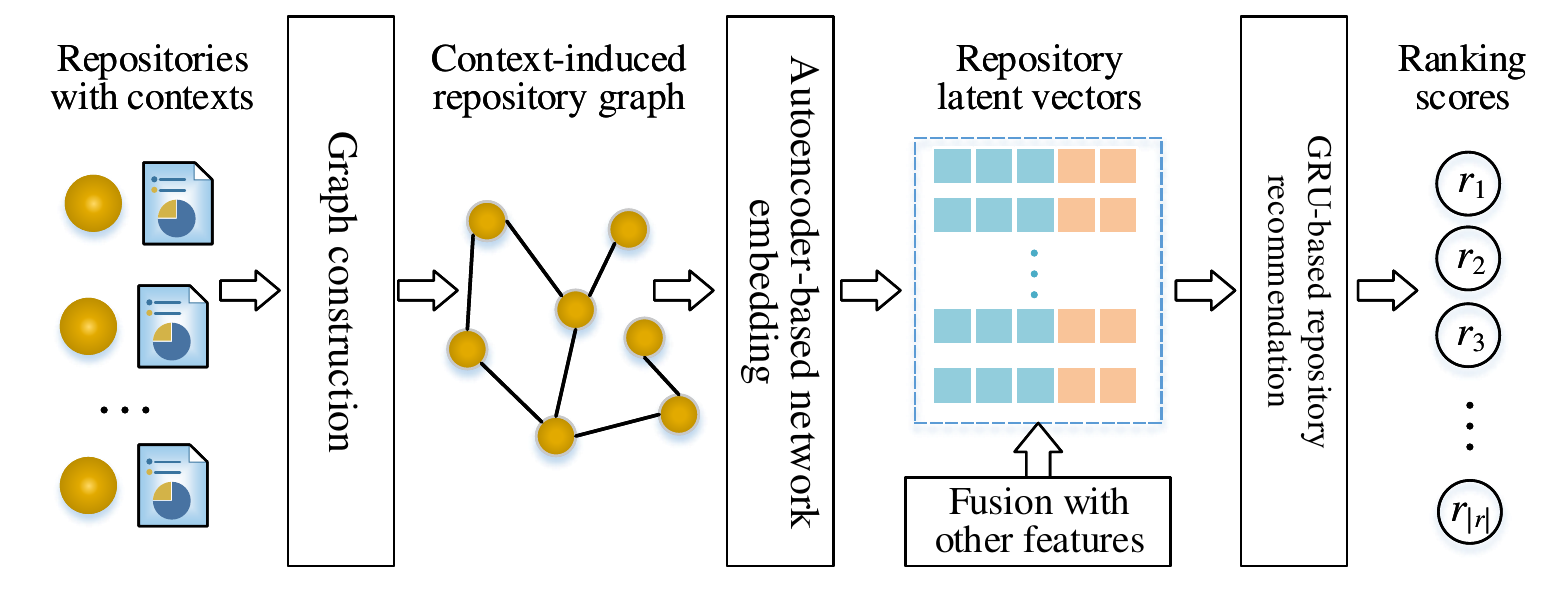}
	\caption{Framework of CSSR.} \label{workflow}
\end{figure}

As illustrated in Fig.~\ref{workflow}, the proposed model named CSSR (Context-aware Sequential Software Service Recommendation) consists of three steps: (1) constructing a repository graph by leveraging contextual information of repositories; (2) feeding the constructed graph into a graph autoencoder to derive embedding of each repository, which is fused with other features to form final repository latent vectors; (3) predicting the probability that each repository will be preferred by each user using a GRU model. These steps are detailed respectively in the following subsections. 

\subsection{Context-Induced Repository Graph Construction} \label{sec/graphconstruction}
First, a repository graph is constructed where the text-based contextual information is exploited. On GitHub, developers usually tag their repositories with topics using words or phrases. 
The topics are suggested by a topic extraction framework, called \emph{repo-topix}, which was developed by GitHub considering many engineering problems. We utilize the topics tagged by users with the suggestion of \emph{repo-topix} directly rather than extracting the similar information using topic modeling like in existing studies. However, some repositories may not be tagged with such information explicitly or tagged incompletely. We then, for each repository, exploit its general description and \textit{R\!E\!A\!D\!M\!E} to derive and complete its topics by techniques such as keyword matching against the explicitly-tagged topics. In addition, the programming language of a repository can also be used as a special kind of topic-like information. This is because users are more likely to find source codes with languages they have used before.

%Even though some repositories are not tagged with such information explicitly, most repositories have general description and \textit{R\!E\!A\!D\!M\!E}, from which we can derive the topics by techniques such as keyword matching against the repositories with explicit topics. 
%The topics describe the purpose of a repository and the type of contents it encapsulates.

Let $\mathcal{T}=\{t_1, t_2, ..., t_{|\mathcal{T}|}\}$ be the set of topics of repositories $\mathcal{R}=\{r_1, r_2, ..., r_{|\mathcal{R}|}\}$. Each repository $r$ has a repository topic vector $\mathcal{RT}=\{rt_1, rt_2, ..., rt_{|\mathcal{T}|}\}$, where $rt_k =1$ if $r$ has $t_k$ (either directly tagged by developers or derived from description, \textit{R\!E\!A\!D\!M\!E}); otherwise $rt_k=0$.

Given any two repositories $r_p$ and $r_q$, $\mathcal{RT}_p= \{rt_{p\cdot1}, rt_{p\cdot2},$ ..., $rt_{p\cdot|\mathcal{T}|}\}$ and $\mathcal{RT}_q=\{rt_{q\cdot1}, rt_{q\cdot2}, ..., rt_{q\cdot|\mathcal{T}|}\}$  are their corresponding repository topic vectors. The similarity between $r_p$ and $r_q$ is measured using cosine distance between $\mathcal{RT}_p$ and $\mathcal{RT}_q$ as follows:  
\begin{equation}
s_{p,q} = \frac{\mathcal{RT}_p \cdot \mathcal{RT}_q}{\|\mathcal{RT}_p\|\|\mathcal{RT}_q\|}
=\frac{\sum_{i=1}^{|\mathcal{T}|}rt_{p\cdot i}\times rt_{q\cdot i}}
{\sqrt{\sum_{i=1}^{|\mathcal{T}|}rt_{p\cdot i}^2}\sqrt{\sum_{i=1}^{|\mathcal{T}|}rt_{q\cdot i}^2}}
\end{equation}

The similarity ranges from 0 (meaning that two repositories don’t have any same topic) to 1 (meaning two repositories have the exactly same set of topics). We further define a hyperparameter in our model, i.e., the edge keeping threshold $\varepsilon\in(0, 1)$, which is used to help generate an effectual and simple graph. With the threshold, the similarity between $r_p$ and $r_q$  is refined as follows:

\begin{equation}\label{similarity}
s_{p,q}=\left\{\begin{array}{ll}
0, & p=q
\\s_{p,q}, &(p\neq q)\wedge(s_{p,q}\geq\varepsilon)
\\0, &(p\neq q)\wedge(s_{p,q}<\varepsilon)
\end{array}\right.
\end{equation}

After calculating the similarity of any two repositories, we can get a similarity matrix $\mathcal{S}=\{\mathbi{s}_1,\mathbi{s}_2, ..., \mathbi{s}_{|\mathcal{R}|}\} \in\mathbb{R}^{|\mathcal{R}|\times|\mathcal{R}|}$, where $\mathbi{s}_r=\{s_{r,1}, s_{r,2}, ..., s_{r,|\mathcal{R}|}\}$. We represent each repository $r$ as a vertex $v_r$. So, there are $\{v_1, v_2, \cdots, v_{|\mathcal{R}|}\}$ vertices; and there is a link between two vertices $v_p$ and $v_q$ only if similarity $s_{p,q}$ is greater than 0. By this way, we obtain a homogeneous graph where the contextual similarity between repositories has been captured. The graph is called \textit{context-induced repository graph}.  

\subsection{Context-Induced Repository Graph Embedding}\label{sec/graphembedding}
\begin{figure}[t]
	\centering
	\includegraphics[width=0.85\textwidth]{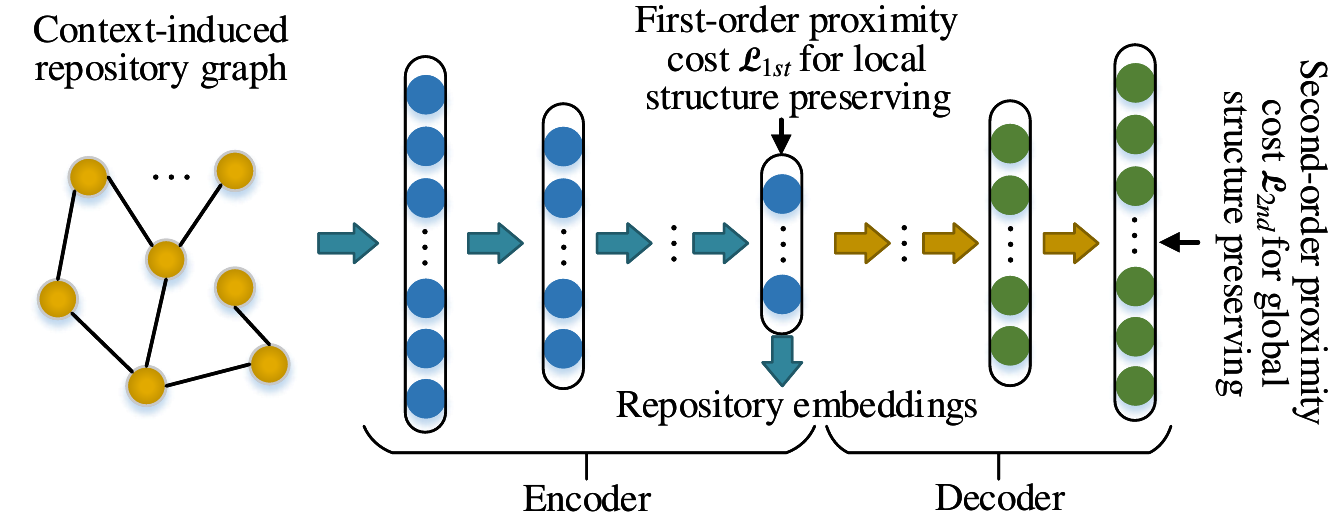}
	\caption{Context-induced repository graph embedding.} \label{GraphEmbedding}
\end{figure}

For nodes in a graph, graph embedding automates the process of extracting low-dimensional node feature vectors. It has been proved very useful in many down-stream tasks, such as classification, link prediction and recommendation. Various graph embedding models have been proposed. For embedding the context-induced repository graph $\mathcal{G}$, we adopt Structural Deep Network Embedding model (SDNE) \cite{SDNE}, a representative embedding model for homogeneous graphs. As illustrated in Fig.~\ref{GraphEmbedding}, a multi-layer autoencoder is extended to capture the non-linear structure of $\mathcal{G}$. The first-order proximity characterizes the local graph structure (i.e., the direct links between repositories in  $\mathcal{G}$). The second-order proximity characterizes the global graph structure (i.e., the co-neighbor relations between repositories in  $\mathcal{G}$). They are jointly exploited in the embedding process.

Given the similarity matrix $\mathcal{S}=\{\mathbi{s}_1,\mathbi{s}_2, ..., \mathbi{s}_{|\mathcal{R}|}\}$ of $\mathcal{G}$, we feed $\mathbi{s}_r=\{s_{r,q}\}_{q=1}^{|\mathcal{R}|}$, $(1\leq r \leq |\mathcal{R}|)$ into an autoencoder. The input $\mathbi{s}_r$ is the initial representation of vertex $v_r$ (i.e., repository $r$). The hidden representations for each layer in the encoder are shown as follows.
\begin{equation}
\begin{array}{l}
\mathbi{y}_r^{(1)}=\sigma(W^{(1)}\mathbi{s}_r + \mathbi{b}^{(1)}),
\\\mathbi{y}_r^{(k)}=\sigma(W^{(k)}\mathbi{y}_r^{(k-1)} + \mathbi{b}^{(k)}), k=2, ..., K
\end{array}
\end{equation}
where $\mathcal{S}=\{\mathbi{s}_r\}_{r=1}^{|\mathcal{R}|}$ is the input data; $Y^{(k)}=\{\mathbi{y}_r^{(k)}\}_{r=1}^{|\mathcal{R}|}$ is the $k$-th layer hidden representations; $W^{(k)}$ and $\mathbi{b}^{(k)}$ are the $k$-th layer weight matrix and biases respectively. $K$ denotes the number of encoder layers. 

Given the input data $\mathbi{s}_r$, we can obtain the final output of the encoder $\mathbi{y}_r^{(K)}$; and then obtain the reconstructed data $\hat{\mathbi{s}}_r$ by reversing the calculation process of encoder (i.e., decoder).     

The first-order proximity describes the pairwise proximity between vertices. Intuitively, it implies that two vertices in real-world graphs are always similar if they are linked by an observed edge. Recall the construction of the context-induced repository graph $\mathcal{G}$ in Sect.~\ref{sec/graphconstruction}, two repositories are certainly similar if they are linked by an observed edge in $\mathcal{G}$. The first-order proximity cost function for local structure preserving is applied on the output layer of the encoder and defined as follows.
\begin{equation}\label{cost1}
\mathcal{L}_{1st}=\sum_{r,q=1}^{|\mathcal{R}|} s_{r,q}\|\mathbi{y}_r^{(K)}-\mathbi{y}_q^{(K)}\|_2^2
\end{equation}
where $s_{r,q}$ is the similarity between any pair of vertices $v_r$ and $v_q$ in $\mathcal{G}$, which is computed by Eq.~(\ref{similarity}).

The second-order proximity exploits the similarity of the vertices’ neighborhood structures to capture the global graph structure. Intuitively, it assumes that if two vertices share many common neighbors, they tend to be similar. The second-order proximity cost function for global structure preserving is applied on the output layer of the decoder. It is shown as follows. 
\begin{equation}\label{cost2}
\mathcal{L}_{2nd}=\sum_{r=1}^{|\mathcal{R}|} \|(\hat{\mathbi{s}}_r-\mathbi{s}_r)\odot\mathbi{b}_r\|_2^2
\end{equation}
where $\odot$ means Hadamard product;  $\mathbi{s}_r$ and $\hat{\mathbi{s}}_r$ are respectively the corresponding input of encoder and the output of decoder for vertex $v_r$ (i.e., repository $r$); $\mathbi{b}_r=\{b_{r,q}\}_{q=1}^{|\mathcal{R}|}$. If $s_{r,q}=0, b_{r,q}=1$; else $b_{r,q}=\beta > 1$. That is, we impose more penalty to the reconstruction error of the non-zero elements than that of zero elements. 
For $\mathbi{s}_r=\{s_{r,q}\}_{q=1}^{|\mathcal{R}|}$ is the direct neighborhood data of repository $r$, the cost function $\mathcal{L}_{2nd}$ ensures the repositories with similar neighborhood to have similar low-dimensional feature vectors. 

We combine Eq.~(\ref{cost1}) and Eq.~(\ref{cost2}) and jointly minimize the following objective function to preserve the first-order and second-order proximity simultaneously:
\begin{equation}
\begin{aligned}
\mathcal{L}_{mix} 
&= \mathcal{L}_{1st} + \alpha\mathcal{L}_{2nd}\\
&=\sum_{r,q=1}^{|\mathcal{R}|} s_{r,q}\|\mathbi{y}_r^{(K)}-\mathbi{y}_q^{(K)}\|_2^2 + \alpha\sum_{r=1}^{|\mathcal{R}|} \|(\hat{\mathbi{s}}_r-\mathbi{s}_r)\odot\mathbi{b}_r\|_2^2
\end{aligned}
\end{equation}

We use the stochastic gradient descent algorithm to train the model. Through the trained model, each repository $r\in\mathcal{R}$ can obtain its initial embedding $\mathbi{y}_r^{(K)}\in\mathbb{R}^{d_r}$, where $d_r$ is the size of graph embeddings.

%where we ignore the regularization term. This is because we use the exact repository latent vectors obtained from graph embedding as initial inputs to make sequential repository recommendation. Following \cite{SDNE}, we use \textit{Deep Belief Network}\cite{DBN} to initialize the parameters and use the stochastic gradient descent algorithm to train the model.

\subsection{Sequential Repository Recommendation}
\begin{figure*}[t]
	\centering
	\includegraphics[width=0.95\textwidth]{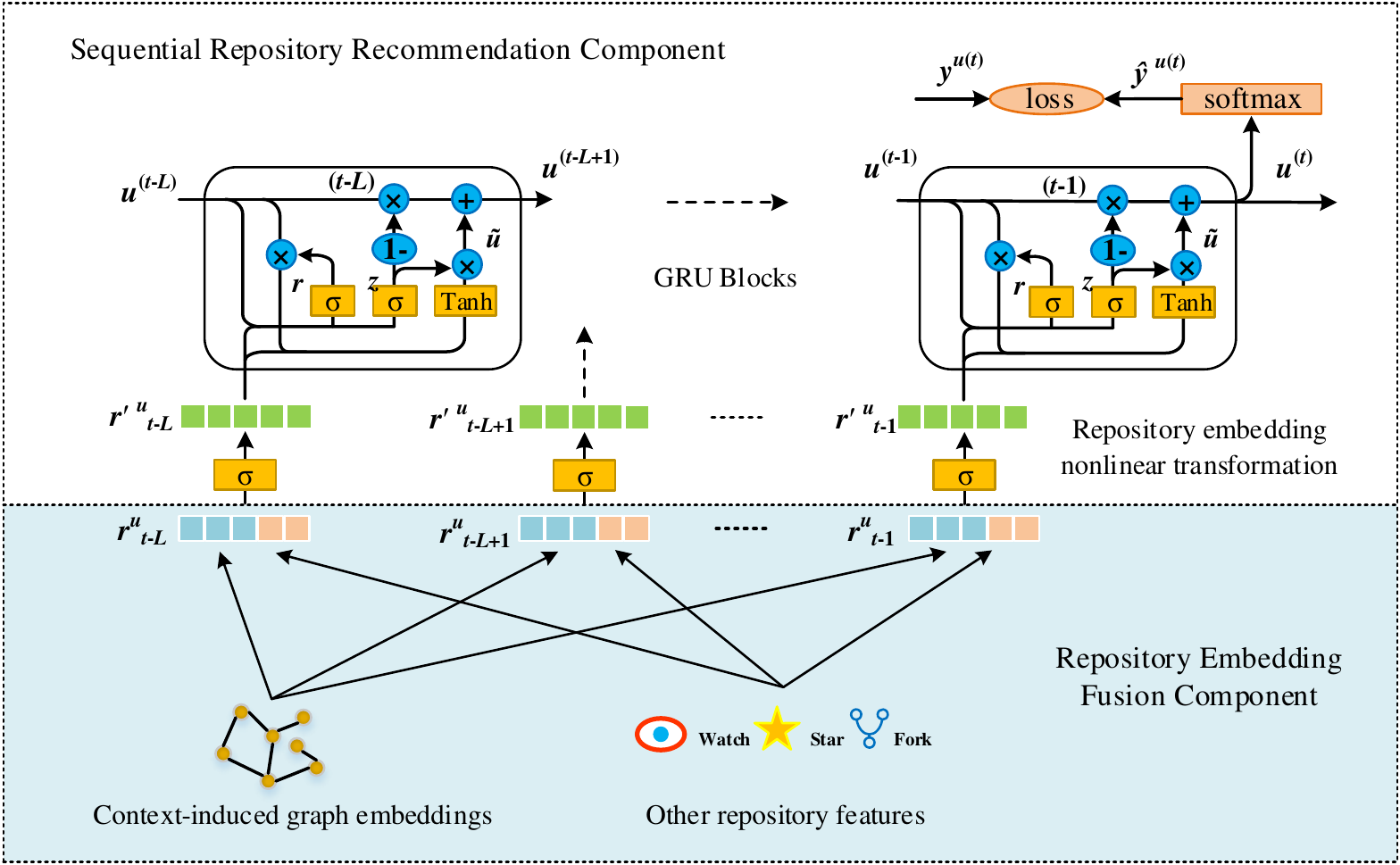}
	\caption{GRU-based repository recommendation.} \label{Framework}
\end{figure*}

The sequential repository recommendation component in the framework of CSSR is illustrated in Fig.~\ref{Framework}. Recall, in a training data record, the features are $R^u_{t-L:t-1}$ and the label is $r^u_t$. More specifically, $R^u_{t-L:t-1}$ includes $L$ repositories, i.e., $r^u_{t-L}$, $\cdots$, $r^u_{t-1}$. Correspondingly, the sequential recommendation component consists of $L$ GRU (Gated Recurrent Unit)\cite{GRU} blocks  as shown in Fig.~\ref{Framework}. The input of GRU block $t-i$ $(1\leq i \leq L)$ is a nonlinear transformed vector $\acute{\mathbi{r}}^u_{t-i}$ of the combined feature vector of repository $r^u_{t-i}$. The repository combined feature vector is derived from graph embeddings as discussed in Sect.~\ref{sec/graphembedding}. It is coupled with a list of other repository features relevant to repository recommendation. The output of the last GRU block is the inferred repository $\hat{y}^{u^{(t)}}$ which will be compared against the ground truth (i.e., $r^u_t$, the label of the training record), and the GRU parameters will be learned. Next, we introduce the sequential repository recommendation modeling in more details.  
\vspace{-3mm}\subsubsection{Fusing Multimodal Features for Repositories.}\vspace{-0.1cm}
For each repository $r_i\in\mathcal{R}  (1\leq i \leq |\mathcal{R}|)$, we concatenate the features derived from context-induced graph embedding with other important features  (i.e., the number of watches, stars and forks) to obtain its final representation. 
Note that these features have different numerical ranges. We adopt min-max scaling to normalize them into a real number between 0 and 1, and concatenate them with the graph embedding vectors. By this way, each repository $r_i^u$ $(t-L \leq i \leq t-1)$ interacted with user $u$ in Fig.~\ref{Framework} can get its final representation (denoted as $\mathbi{r}_i^u\in\mathbb{R}^{d_r+3}$) for a GRU model, where $d_r$ is the dimensionality of initial repository embeddings. 

\vspace{-3mm}\subsubsection{Generating User Representation.}\vspace{-0.1cm}
Next, we present how to derive the latent vector of user $u$ at time step $t$ via the GRU model based on repositories $u$ interacted with before $t$. 

Recall $\mathcal{R}_{t-L:t-1}^u=\{r_{t-L}^u, r_{t-L+1}^u, ..., r_{t-1}^u\}$ is the sequence of repositories of size $L$ interacted with $u$  just before time step $t$. The user representation can be computed iteratively as follows.
\begin{equation}
\begin{array}{l}\label{eq/GRU}
\acute{\mathbi{r}}_i^u = \sigma(W_{in}\cdot \mathbi{r}_i^u),\\
\mathbi{z}^{(i)} = \sigma(W_z[\mathbi{u}^{(i)}, \acute{\mathbi{r}}_i^u]),\\
\mathbi{r}^{(i)} = \sigma(W_r[\mathbi{u}^{(i)}, \acute{\mathbi{r}}_i^u]),\\
\tilde{\mathbi{u}}^{(i+1)} = \tanh(W_u[\mathbi{r}^{(i)}\odot\mathbi{u}^{(i)}, \acute{\mathbi{r}}_i^u]),\\
\mathbi{u}^{(i+1)} =  \mathbi{z}^{(i)}\odot\mathbi{u}^{(i)} + (1-\mathbi{z}^{(i)})\odot\tilde{\mathbi{u}}^{(i+1)},\\
s.t. \hspace{0.3cm} t-L \leq i\leq t-1.
\end{array}
\end{equation}
where  $W_{in}\in \mathbb{R}^{d_u\times(d_r+3)}$, $W_z, W_r, W_u \in \mathbb{R}^{d_u\times 2d_u}$ are learnable parameters, and $d_u$ is the dimensionality of user embeddings. $\sigma(\cdot)$ is a logistic sigmoid function to do non-linear projection. $\odot$ is the Hadamard product between two vectors. $\mathbi{z}^{(i)}$ and  $\mathbi{r}^{(i)}$ denote update gate and reset gate respectively. They control how information flows through the sequence. $\mathbi{u}^{(i)}$ represents the hidden state vector, i.e., the representation of the user $u$ at time step $i$. It is dynamic and can remember past states. $\mathbi{r}_i^u$ is the combined feature vector of the repository interacted with $u$ at time step $i$. $\acute{\mathbi{r}}_i^u$ is its nonlinear transformed vector, here, as the input data of the $i$-th GRU block. Note that the dimensionalities of $\mathbi{u}^{(i)}$ and $\acute{\mathbi{r}}_i^u$ are the same (i.e., $d_u$) for the convenience of prediction, and $\mathbi{u}^{(t-L)}=\vec{\mathbf{0}}$ when initializing. For simplicity, we use  $\mathbi{u}^{(i+1)}=\rm{GRU}(\mathbi{u}^{(i)}, \mathbi{r}_i^u)$ to denote the above operations.

\vspace{-3mm}\subsubsection{Recommending and Model Training.}
We then illustrate how to recommend repositories for $u$ at $t$. Given a user $u$ with a sequence of visited repositories $\mathcal{R}_{t-L:t-1}^u$, the embedding $\mathbi{u}^{(t)}$ of user $u$ at time step $t$ can be computed by Eq.~(\ref{eq/GRU}). Then, the rating score of each candidate repository $r_i\in\mathcal{R}$   $(1\leq i \leq |\mathcal{R}|)$ for user $u$ is computed as follows.
\begin{equation}
\hat{y}_i^{u^{(t)}} = {\rm softmax}( \mathbi{u}^{(t)} \cdot \acute{\mathbi{r}}_i )
\end{equation}
where $\hat{y}_i^{u^{(t)}}$ denotes the recommendation probability of repository $r_i$ to user $u$ at time step $t$. Finally, we train the GRU model by minimizing the following objective function.
\begin{equation}
\begin{aligned}
\mathcal{L} 
&= \mathcal{L}_{cross-entropy} + \mathcal{L}_{reg}\\
&= -\sum_{u\in \mathcal{U}}\sum_{t\in \mathcal{T}_u}\sum_{i\in \mathcal{T}_u^t} y_i^{u^{(t)}}\log(\hat{y}_i^{u^{(t)}}) + \lambda\|\theta\|_2^2
\end{aligned}
\end{equation}
where $\mathcal{T}_u$ denotes the training dataset of user $u$, and  $\mathcal{T}_u^t\subset \mathcal{R}$ denotes the training set of user $u$ at time step $t$. We adopt the negative sampling strategy to train the model. $\mathcal{T}_u^t$  contains the repository $r_u^t$ that user $u$ interacted with at time step $t$ and the randomly-sampled 10 negative repositories corresponding to $r_u^t$. $y_i^{u^{(t)}}$ denotes the ground truth. If repository $r_i = r_u^t$, $y_i^{u^{(t)}}=1$. Otherwise, $y_i^{u^{(t)}}=0$. $\mathcal{L}_{reg}$ is an L2-norm regularizer to prevent overfitting, and $\theta$ is the set of all learnable parameters. 

\section{Experiments and Evaluation}\label{sec/experiments}
In this section, we conduct experiments to validate the effectiveness of the proposed model in capturing dynamics of user preferences and addressing data sparsity issue. We also analyze how the key hyperparameters affect the performance.

\subsection{Dataset Description}
We evaluate the proposed method on a large dataset crawled from GitHub. We use GitHub REST API to create calls to obtain the data in JSON format and store them in MongoDB. A user is considered to prefer a repository if the user forked it. This means the user produced a personal copy of someone else's repository so that she can contribute to it or use it as the starting point for her own. All the repositories that a user created or forked are listed in her Web page in a chronological order. We randomly select users who forked more than 5 repositories. For each selected user, we crawl the information of all her forked repositories (e.g. topics, programming languages, \textit{R\!E\!A\!D\!M\!E}). Such information is used to construct context-induced repository graph. We eliminate repositories forked by fewer than 5 users. There are 2,616 users, 3,126 repositories, and 21,924 interactions in the dataset after preprocessing. The data sparsity is 0.268\%.
 
To capture dynamics of user preferences, we also try to consider temporal information of user-repository interactions (e.g., the absolute time span, the relative time interval, the relative temporal position interval) in our model. However, the experimental results don't show performance improvement. Thus, CSSR just utilizes the sequential interactions to make repository recommendation as described above.

\subsection{Baseline Methods}
Next, we compare CSSR against the following baseline methods.
\begin{itemize}
	\item \textbf{Pop} simply recommends top ranked repositories based on popularity in training data.
	\item  \textbf{Item-KNN} \cite{Item_KNN} recommends a user the repositories similar to the previously forked repositories by the user based on cosine similarity.	
	\item \textbf{BPR} \cite{BPR} is a classic method for non-sequential recommendation, which optimizes a Matrix Factorization model using a pairwise ranking loss. 
	\item \textbf{FFM} \cite{FFM} is the representative recommendation model based on factorization machine. It groups features into fields, and learns the interactions between users and repositories to complete the user-repository implicit rating matrix.
	\item \textbf{GRU4Rec} \cite{GRU4Rec} is a representative sequential recommendation model, which also utilizes GRU to	model user action sequences. We feed randomly-generated repository embeddings into GRU blocks, and obtain the best performance by using Xavier initializer against other random number generators. 	 
	\item \textbf{PNCF} \cite{PNCF} is the state-of-the-art GitHub repository recommendation method by building a  preference-based neural collaborative filtering recommender model. We feed the model not only with the language features in the original paper but also all our utilized topic features to make it fair.
\end{itemize}

\subsection{Experimental Settings}
For constructing the context-induced repository graph, we conduct stemming and lemmatization on all topics tagged by users with the help of \emph{repo-topix}, and extract 4,015 topics. Then, for each repository, we conduct tokenization, stop words removing, stemming, lemmatization, typo corrections, and string matching on its textual contexts to complete its topics as introduced in Sect.~\ref{sec/graphconstruction}. RE, NLTK and Spacy packages of Python are used for these tasks. Next, we compute the cosine similarity between each pair of repositories by using Numpy package and save the constructed graph into a file by using Pandas package. The edge keeping threshold $\varepsilon$ is set to 0.3, which can generate a reasonable number of edges. The final constructed graph contains 3,126 repositories with 168,039 edges between them. 

For each user, we hold the first 80\% of interactions as the training set. We then use the next 10\% of interactions as the validation set for hyperparameter tuning. The latest 10\% constitute the test set for reporting model performance. We conduct experiments with each model for five times independently, and report the average results. 

The hyperparameters are learned from the validation dataset and set as follows. The size of initial repository embedding is 140. The size of user embedding is 64. $L=4$ means that the recent 4 interacted repositories are considered to infer recommendations. The learning rate is 0.009. The maximum number of epoch is 100 during the model training. All the experimental results of our model are achieved by using the above hyperparameter configuration settings if no specific situations are provided. The optimal hyperparameters of each baseline method are set based on the experiment reports of the relevant research papers. We implement our CSSR model in Tensorflow.

We adopt three commonly-used metrics to evaluate the recommendation performance, i.e., Hit Rate (\textit{HR@N}), Mean Reciprocal Rank (\textit{MRR@N}) and Normalized Discounted Cumulative Gain (\textit{NDCG@N}) where $N$ indicates top-$N$ ranked repositories. In general, \textit{HR@N} doesn't care about rank position in the recommendation list, \textit{MRR@N} considers only the position of the first matched recommendation, and \textit{NDCG@N} is a full position-aware metric which assigns greater weights on higher positions. They reflect different aspects of recommendation quality. The higher values indicate the better repository recommendation quality for all the three metrics. 

\subsection{Performance Comparison}
\begin{table*}[t]
	\caption{The performance comparison (The method with the best performance is starred and the method with the second-best performance  is boldfaced; columns ``KNN'' and ``GRU'' denote the baseline ``Item-KNN'' and ``GRU4Rec'' respectively; column ``Improv.'' denotes the improvement ratio of CSSR relative to the best baseline).} \label{comparison}
	\footnotesize
	\begin{tabular}{p{0.95cm}<{\centering}|p{1.7cm}|p{1.05cm}<{\centering}|p{1.05cm}<{\centering}|p{1.05cm}<{\centering}|p{1.05cm}<{\centering}|p{1.05cm}<{\centering}|p{1.05cm}<{\centering}|p{1.05cm}<{\centering}||p{1.05cm}<{\centering}}
		\hline
		top-N                & Metrics     & Pop  & KNN        & BPR   & FFM   & GRU   & PNCF           & CSSR       & Improv. \\
		\hline
		& HR   (\%)   & 1.566 & 2.467 & 3.103 & 2.709 &3.086 & \textbf{3.231}          & \textbf{3.769}* & 16.65\% \\
		& MRR   (\%)  & 0.662 & 1.421 & 1.646 & 1.428 &1.288 & \textbf{1.705}          & \textbf{2.085}* & 22.29\% \\
		\multirow{-3}{*}{5}  & NDCG   (\%) & 0.886 & 1.923 & 2.003 & 1.746 &1.712 & \textbf{2.079}          & \textbf{2.497}* & 20.11\% \\
		\hline
		& HR   (\%)   & 2.428 & 3.955 & 4.713 & 3.926 &4.414 & \textbf{4.962}          & \textbf{6.077}* & 22.47\% \\
		& MRR   (\%)  & 0.762 & 1.628 & 1.857 & 1.586 &1.448 & \textbf{1.926}          & \textbf{2.378}* & 23.47\% \\
		\multirow{-3}{*}{10} & NDCG   (\%) & 1.151 & 2.599 & 2.519 & 2.136 &2.099 & \textbf{2.629}          & \textbf{3.206}* & 21.95\% \\
		\hline
		& HR   (\%)   & 4.855 & 5.090          & \textbf{6.206} & 5.300 &5.781 & 6.115 & \textbf{7.308}* & 17.76\% \\
		& MRR   (\%)  & 0.962 & 1.723 & 1.975 & 1.694 &1.576 & \textbf{2.018}          & \textbf{2.472}* & 22.50\% \\
		\multirow{-3}{*}{15} & NDCG   (\%) & 1.803 & \textbf{3.064} & 2.915 & 2.499 &2.501 & 2.935          & \textbf{3.523}* & 14.98\% \\
		\hline
		& HR   (\%)   & 5.834 & 5.991          & \textbf{7.384} & 6.203 &6.680 & 7.269 & \textbf{8.577}* & 16.16\% \\
		& MRR   (\%)  & 1.016 & 1.774 & 2.042 & 1.745 &1.620 & \textbf{2.082}          & \textbf{2.541}* & 22.05\% \\
		\multirow{-3}{*}{20} & NDCG   (\%) & 2.032 & \textbf{3.444} & 3.194 & 2.712 &2.685 & 3.206          & \textbf{3.835}* & 11.35\%\\
		\hline
	\end{tabular}
\end{table*}

\begin{table*}[t]
	\caption{The performance comparison at different sparsity levels.}\label{sparsity}
	\footnotesize
	\begin{tabular}{p{1.35cm}<{\centering}|p{1.7cm}|p{1cm}<{\centering}|p{1cm}<{\centering}|p{1cm}<{\centering}|p{1cm}<{\centering}|p{1cm}<{\centering}|p{1cm}<{\centering}|p{1cm}<{\centering}||p{1cm}<{\centering}}
		\hline
		\tabincell{c}{Ratio\\(Sparsity)}                & Metrics     & Pop  & KNN        & BPR   & FFM  & GRU & PNCF           & CSSR       & Improv. \\
		\hline
		& HR   (\%)   & \textbf{2.346} &	1.006 &	2.194 &	1.735 &1.904 &	2.250 &	\textbf{3.369}* & 43.61\% \\
		& MRR   (\%)  & 0.741&	0.402&	0.748&	\textbf{0.927}&0.711 &	0.772&	\textbf{1.457}*&	57.17\%\\
		\multirow{-3}{*}{\tabincell{c}{ALL\\(0.096\%)}} & NDCG   (\%) & 1.103&	0.543&	1.085&	1.116 &0.989 &	\textbf{1.119}&	\textbf{1.881}*&	68.09\%\\
		\hline
		& HR   (\%)   & 2.747&	2.709&	3.403&	2.624 &3.633 &	\textbf{3.640}&	\textbf{4.648}*&	27.69\%\\
		& MRR   (\%)  & 0.924&	0.872&	1.434&	1.067 &1.323 &	\textbf{1.511}&	\textbf{1.922}*&	27.20\%\\
		\multirow{-3}{*}{\tabincell{c}{Half\\(0.182\%)}}  & NDCG   (\%) & 1.350&	1.538&	1.893&	1.429 &1.853 &	\textbf{2.004}&	\textbf{2.487}*&	24.10\%\\
		\hline
		& HR   (\%)   & 2.428 & 3.955 & 4.713 & 3.926 &4.414 & \textbf{4.962}          & \textbf{6.077}* & 22.47\% \\
		& MRR   (\%)  & 0.762 & 1.628 & 1.857 & 1.586 &1.448 & \textbf{1.926}          & \textbf{2.378}* & 23.47\% \\
		\multirow{-3}{*}{\tabincell{c}{No\\(0.268\%)}} & NDCG   (\%) & 1.151 & 2.599 & 2.519 & 2.136 &2.099 & \textbf{2.629}          & \textbf{3.206}* & 21.95\% \\
		\hline
	\end{tabular}
\end{table*}

The experimental results of of CSSR and all the baselines are reported in Table~\ref{comparison}. We have the following observations.
(1) In most cases, the state-of-the-art baseline PNCF achieves the best performance than the other baseline methods.
(2) The proposed CSSR consistently achieves better performance on all the metrics at different $N$ values compared with all the baselines by at least 10\%. Specifically, it improves the performance slightly more on the metric MRR than on HR and NDCG. It achieves slightly more performance improvement when $N=10$ than other $N$ values. 
(3) All the baselines except GRU4Rec are sequential-information free models. However, GRU4Rec doesn't outperform the other baselines by just using randomly-initialized repository embeddings. Compared with GRU4Rec, the significant improvement of CSSR validates the importance of the context-induced repository graph embedding component in our model.

\subsection{Impact of Data Sparsity}
We compare CSSR and all the baselines at different levels of data sparsity. The aim is to evaluate the solution applied in CSSR for mitigating the issue of sparse data. Since each user has at least 3 repositories and a repository has at least one user, we delete at most 14,081 interactions in our dataset to simulate different settings of sparsity. Table~\ref{sparsity} shows the performance of all the methods at three levels of sparsity, i.e., deleting all/half of/none of the 14081 interactions respectively. We set $N$=10, and adopt all repositories in the training set that a user has interacted with to train the model in the first two sparsity levels, i.e., the repository sequence length $L$ is not fixed. From Table~\ref{sparsity}, we have the following observations.
(1) The performance of all the methods gets worse and worse when the data sparsity changes from 0.268\% to 0.182\% and then to 0.096\%. 
(2) Compared with all the baselines except Pop, the impact of data sparsity on CSSR is much weaker.
CSSR can have much more stable performance than the other methods except Pop, and achieve more significant improvements against the best baseline in the sparser data set. 
(3) Although the impact of data sparsity on Pop is weaker than CSSR, the performance of Pop is very poor among baselines. In short, our model achieves the best performance and demonstrates robustness in the situation of data sparsity.

\subsection{Sensitivity of Hyperparameters}
\begin{figure*}[t]
	\includegraphics[width=\textwidth]{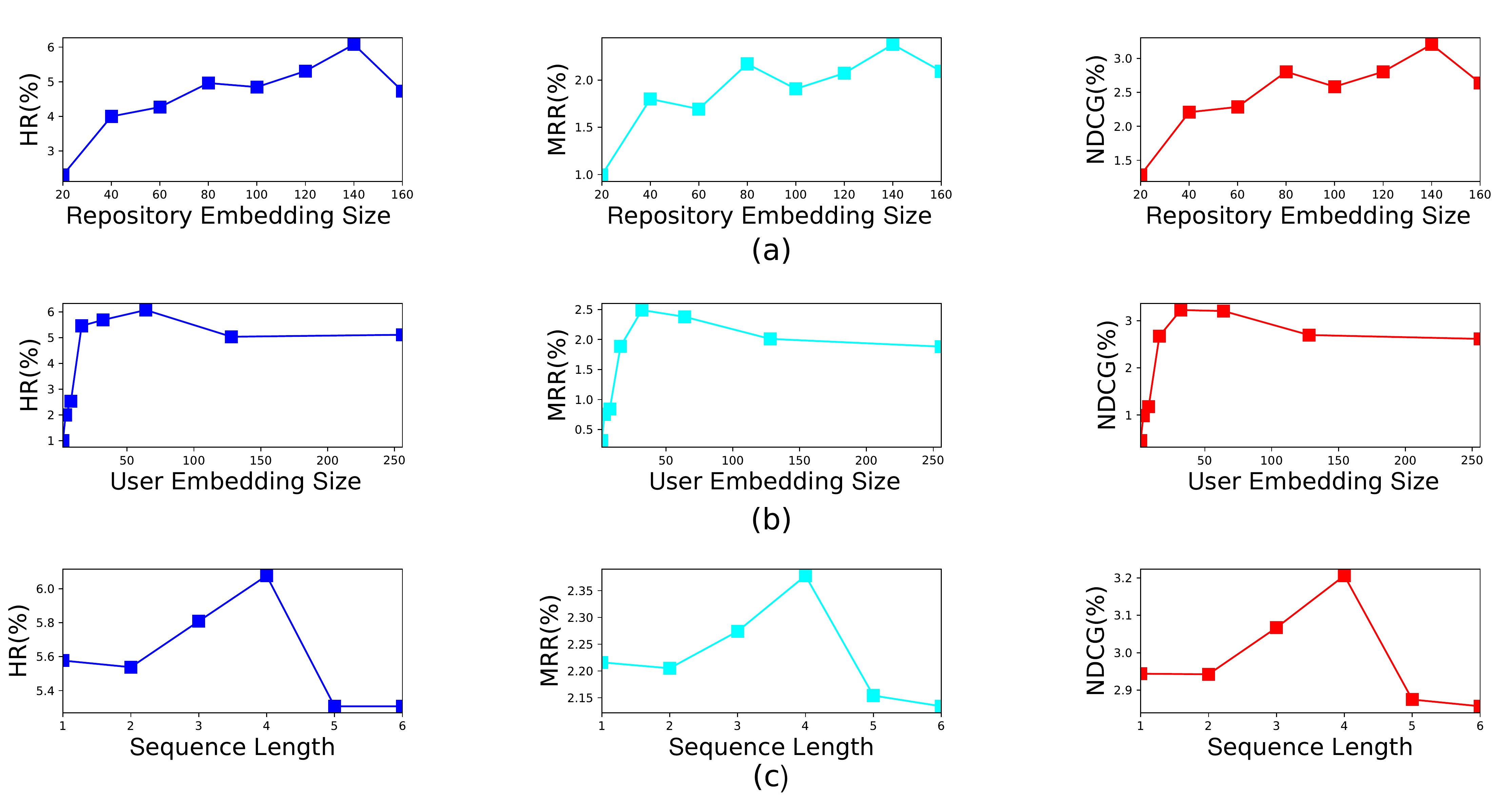}
	\caption{The sensitivity of CSSR performance to hyperparameters (a) repository embedding size; (b) user embedding size; (c) sequence length ($L$).} \label{effects}
\end{figure*}
The size of initial repository embeddings, the size of user embeddings, and the sequence length $L$ are the three important hyperparameters in our model. In this study, we further investigate the impact of the three hyperparameters on the recommendation performance. We set $N$=10.
Fig.~\ref{effects} (a) illustrates the effect of the initial repository embedding size (i.e., the size of graph embeddings). We can see that our CSSR would achieve the best performance when the graph embedding size is 140. Fig.~\ref{effects} (b) illustrates the effect of the user embedding size (i.e., the size of hidden state in the GRU model).  We can see that CSSR would achieve the best performance when the user embedding size is 32 or 64. Fig.~\ref{effects} (c) illustrates the effect of the sequence length $L$. We can see that CSSR would achieve the best performance when it equals 4.

\section{Conclusion}\label{sec/conclusion}
This paper presented a context-aware sequential software service recommendation model---CSSR. It can recommend repositories on GitHub matching users' interests. CSSR is a joint model that incorporates a graph embedding technique into a GRU formulation to generate latent vectors of users and repositories. Specifically, graph embedding technique is leveraged to exploit rich repository contextual information to alleviate the data sparsity problem. The context-aware latent vectors of repositories are then fed into a GRU model, which captures the dynamics of user preference and eventually recommend repositories to users. 
The results of extensive experiments show that our method can significantly outperform the existing state-of-the-art repository recommender models in various aspects.

%In the future, we will attempt to leverage more deep sequential neural models like Transformer to improve the performance of CSSR further.  Moreover, we will conduct studies on how to recommend suitable repositories matching a Web service or a mashup specified by users.

\vspace{3mm}\noindent\small\textbf{Acknowledgements.} This work is partially supported by Australian Research Council Linkage Project (No.LP180100750) and Discovery Project (No.DP210100743).
\vspace{-3mm}


\bibliographystyle{splncs04}
\bibliography{references}

\end{document}